# A Soluble Modified Fermi-Hubbard Model

Moorad Alexanian

*Department of Physics and Physical Oceanography*
*University of North Carolina Wilmington, Wilmington, NC 28403-5606*

E-mail: alexanian@uncw.edu

**Abstract.** A recently introduced recurrence-relation ansatz applied to the Bose-Hubbard model is here used in the Fermi-Hubbard model. The resulting modified Fermi-Hubbard model is soluble and exhibits a continuous phase transition (second order) reminiscent of the integer quantum Hall resistance and a ground-state, first-order phase transition.

## 1. Introduction

The Fermi-Hubbard model has become the basis for much of what we know about superfluidity in periodic systems, quantum magnetism, and strongly correlated fermion physics on lattices in general [1]. Here, we consider a recently introduced recurrence-relation ansatz between annihilation operators in the Bose-Hubbard model [2] and apply it to the hopping amplitude of the Fermi-Hubbard model. The ansatz considerably simplifies the Fermi-Hubbard model to the extent that the resulting modified model is exactly soluble.

This paper is structured as follows. In Sec. 2, we present the Fermi-Hubbard model on an infinite, one-dimensional lattice. In Sec. 3, we introduce a recurrence-relation ansatz for the external degree of freedom associated with the nearest neighbor of the $j$-th lattice site. In Sec. 4, the grand canonical partition function is determined. In Sec. 5, the average number on microstates in a particular lattice site is determined. In Sec. 6, the on-site density fluctuations and sites correlations are found. In Sec. 7, we obtain both the energy and the entropy of the system. Finally, Sec. 8, we summarizes our results.

## 2. Fermi-Hubbard model

The SU(N) Fermi-Hubbard model is given by [3, 4]

$$\hat{H} = -t \sum_{\langle i,j \rangle, \sigma} (\hat{c}^\dagger_{i\sigma} \hat{c}_{j\sigma} + h.c.) + \frac{U}{2} \sum_{i, \sigma \neq \tau} \hat{n}_{i\sigma} \hat{n}_{i\tau} - \sum_{i,\sigma} \mu_i \hat{n}_{i\sigma}, \qquad (1)$$

where $\hat{c}^\dagger_{i\sigma}$ and $\hat{c}_{i\sigma}$ represent the fermionic creation and annihilation operators at site $i$ with $\sigma \in \{1...N\}$, $\hat{n}_{i\sigma} = \hat{c}^\dagger_{i\sigma} \hat{c}_{i\sigma}$ spin is the number operator, $\langle i, j \rangle$ denotes adjacent sites on a rectangular lattice, $t$ is the hopping amplitude, $U$ is the on-site interaction strength, and $\mu_i$ denotes the chemical potential. Representing the Fermi-Hubbard Hamiltonian on a quantum computer requires a fermionic encoding. The well-known Jordan-Wigner transform is used, under which each fermionic mode maps to one qubit, interpreted as lying on a 1D line [3].

## 3. Ansatz

Consider the following recurrence-relation ansatz for the term associated with the hopping term of the *i*-th lattice in (1)

$$\hat{c}_{i+1\sigma} = \hat{c}_{i\sigma} - \hat{c}_{i-1\sigma}. \tag{2}$$

and so

$$\sum_{i,\sigma}\left[\hat{c}^\dagger_{i\sigma}\hat{c}_{i+1\sigma} + \hat{c}^\dagger_{i+1\sigma}\hat{c}_{i\sigma} + \hat{c}^\dagger_{i\sigma}\hat{c}_{i-1\sigma} + \hat{c}^\dagger_{i-1\sigma}\hat{c}_{i\sigma}\right] = 2\sum_{i,\sigma}\hat{c}^\dagger_{i\sigma}\hat{c}_{i\sigma} = 2\sum_i \hat{n}_i, \tag{3}$$

$$\hat{n}_i = \sum_\sigma \hat{n}_{i\sigma}.$$

where

The Fermi-Hubbard model (1) is reduced to

$$\begin{aligned}\hat{H} &= -2t\sum_i \hat{n}_i + \frac{U}{2}\sum_{i,\sigma\neq\tau}\hat{n}_{i\sigma}\hat{n}_{i\tau} - \sum_{i,\sigma}\mu_i\hat{n}_{i\sigma},\\ &= -2t\sum_i \hat{n}_i + \frac{U}{2}\sum_i \hat{n}_i^2 - \frac{U}{2}\sum_i \hat{n}_i - \sum_i \mu_i \hat{n}_i,\end{aligned} \tag{4}$$

since $\hat{n}_{i\sigma}^2 = \hat{n}_{i\sigma}$ with energy eigenvalues

$$E_i = -2tn_i + \frac{U}{2}n_i^2 - \frac{U}{2}n_i - \mu_i n_i. \tag{5}$$

## 4. Grand canonical partition function

The grand canonical partition function is the following product over the differing lattices

$$\mathscr{Z} = \prod_{i,\sigma} e^{-\beta E_i}, \tag{6}$$

where $\beta = 1/k_B T$ and we consider $N$ lattice sites with each containing up to $N$ distinguishable microstates $\sigma$, viz., $\sum_i n_i = N$ and so $\sum_i \langle \hat{n}_i \rangle = N$. One obtains, with the aid of (5),

$$\mathscr{Z} = \prod_i \sum_{n=0}^N \frac{N!}{n!(N-n)!} e^{\left[(2t/U+1/2+\mu_i/U)n - n^2/2\right]U/(k_BT)} = \prod_i \sum_{n=0}^N \frac{N!}{n!(N-n)!} e^{(\tilde{\mu}_i n - n^2/2)/\tilde{T}}, \tag{7}$$

where the renormalized, scaled chemical potential $\tilde{\mu}_i$ and the scaled temperature $\tilde{T}$ are given by

$$\tilde{\mu}_i = 2t/U + 1/2 + \mu_i/U \quad \text{and} \quad \tilde{T} = k_B T/U \tag{8}$$

and the binomial coefficient $\frac{N!}{n!(N-n)!}$ represents the distribution of $n$, $n = 0 \cdots N$, distinguishable microstates in the *i*-th, $i = 1 \ldots N$, distinguishable lattice sites with all parameters expressed in units of $U$. Accordingly, there are only the variables $N$, $\tilde{\mu}_i$ for each lattice site $i$, and $\tilde{T}$.

It is important to remark that a given value of the renormalized, scaled chemical potential $\tilde{\mu}_i$ does not determine the individual values of either *t/U* or $\mu_i/U$. It is clear that our modified Fermi-Hubbard model reduces to the original Fermi-Hubbard model for *t = 0*. It may be that the results from the modified Fermi-Hubbard model for *t > 0* and $\mu_i \gg t$ reproduce those of the original Fermi-Hubbard model. In what follows, we calculate various thermodynamic variables for the modified Fermi-Hubbard model given by the grand partition function (7).

## 5. Average number

The average number $\langle \hat{n} \rangle$ of microstates in a lattice site with chemical potential $\tilde{\mu}$ and temperature $\tilde{T}$ is given by

$$n_{ave} \equiv \langle \hat{n} \rangle = k_B T \frac{\partial \ln \mathscr{Z}}{\partial \mu} = \frac{\sum_{n=0}^{N} \frac{N!}{n!(N-n)!} n e^{(\tilde{\mu}n - n^2/2)/\tilde{T}}}{\sum_{n=0}^{N} \frac{N!}{n!(N-n)!} e^{(\tilde{\mu}n - n^2/2)/\tilde{T}}}. \tag{9}$$

Figs.1-3 show the behavior of the average number of microstates in a lattice site for particular values of the temperature $\tilde{T}$ as a function of the chemical potential $\tilde{\mu}$. The asymptotic behavior of $\langle \hat{n} \rangle$ for large $\tilde{\mu}$ requires $\tilde{\mu} \gg N - 1/2$, which at low temperatures $\tilde{T}$ may be reached at smaller values of the chemical potential $\tilde{\mu}$, viz., $\tilde{\mu} \approx N$, as illustrated in Figs.1-3. The stepwise asymptotic behavior in Figs.1-3 represent actually a continuous phase transition (second order) as indicated in the temperature dependence in Figs.1-3, viz., as the temperature increases, the steps vanish while the asymptotic behavior remains. This behavior is reminiscent of the integer quantum Hall resistance [5, 6].

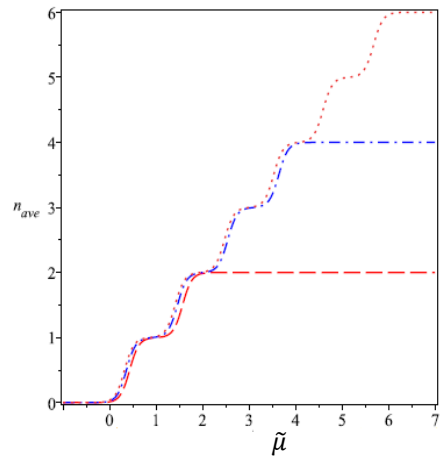

**Fig. 1.** Average number of microstates in a given lattice site versus the chemical potential $\tilde{\mu}$ for $\tilde{T}$=0.10. Red *N* = 2, blue *N* = 4, and orange *N* = 6.

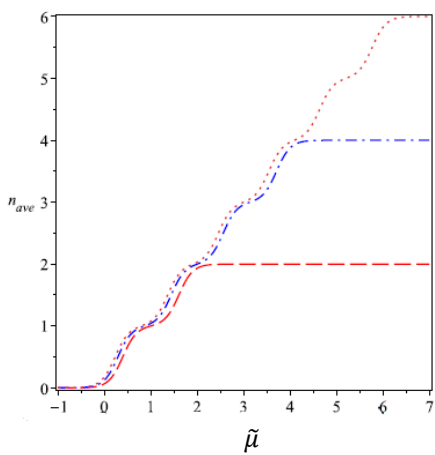
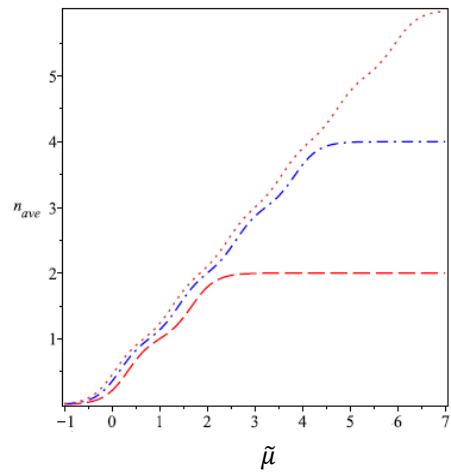

**Fig. 2.** Same as Fig.1 but with $\tilde{T}$=0.15.     **Fig. 3.** Same as Fig.1 but with $\tilde{T}$=0.25.

## 6. On-site density fluctuations and correlations

The on-site density fluctuations is

$$\delta n^2 = \langle \hat{n}^2 \rangle - \langle \hat{n} \rangle^2 = \frac{\sum_{n=0}^{N} \frac{N!}{n!(N-n)!} n^2 e^{(\tilde{\mu}n - n^2/2)/\tilde{T}}}{\sum_{n=0}^{N} \frac{N!}{n!(N-n)!} e^{(\tilde{\mu}n - n^2/2)/\tilde{T}}} - \left( \frac{\sum_{n=0}^{N} \frac{N!}{n!(N-n)!} n e^{(\tilde{\mu}n - n^2/2)/\tilde{T}}}{\sum_{n=0}^{N} \frac{N!}{n!(N-n)!} e^{(\tilde{\mu}n - n^2/2)/\tilde{T}}} \right)^2. \quad (10)$$

Fig.4 shows the behavior of the on-site density fluctuations for $N = 6$ as a function of $\tilde{\mu}$ for increasing values of the temperature $\tilde{T}$. The peaks in Fig.4 at low temperatures are associated with the transition from one step to the next of Figs.1-3. A number of $N$ peaks occur for any value of $N$ and the behavior is similar as the temperature increases and the system goes through the continuous phase transition (second order) as the peaks disappear. The isothermal compressibility is defied by

$$\kappa_T \equiv \left( \frac{\partial \langle \hat{n} \rangle}{\partial \mu} \right)_T = T^{-1}(\langle \hat{n}^2 \rangle - \langle \hat{n} \rangle^2) \quad (11)$$

and so the behavior of the isothermal compressibility is essentially represented also by Fig.4. Figs.5-6 show the on-site density fluctuations but now as a function of $\tilde{T}$. It is interesting that in the limit $\tilde{T} \to \infty$, the density fluctuations approaches the value of 3/2, which for general $N$ is given by $N/4$.

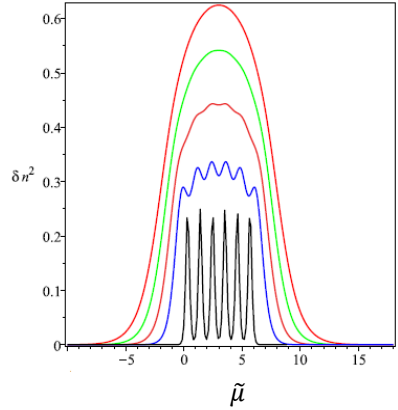

**Fig. 4.** On-site density fluctuations $\delta n^2$ versus renormalized scaled $\tilde{\mu}$ for N = 6 with $\tilde{T} = 1.0$ (red), $\tilde{T} = 0.8$ (green), $\tilde{T} = 0.6$ (orange), $\tilde{T} = 0.4$ (blue), and $\tilde{T} = 0.1$ (black). The fluctuation $\delta n^2 \to 0$ for $|\mu| \to \infty$ for $\tilde{T} < \infty$. The isothermal compressibility is $T\kappa_T = \delta n^2$.

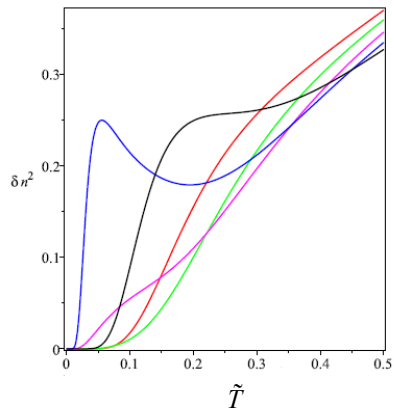

**Fig. 5.** On-site density fluctuations $\delta n^2$ versus scaled temperature $\tilde{T}$ for $N = 6$ with $\tilde{\mu} = 1.0$ (red), $\tilde{\mu} = 0.8$ (green),

$\tilde{\mu} = 0.6$ (magenta), $\tilde{\mu} = 0.4$ (blue), and $\tilde{\mu} = 0.1$ (black). The correlations vanish at $\tilde{T} = 0$ with an essential singularity and approaches the value of $N/4$ as $\tilde{T} \to \infty$ for $|\tilde{\mu}| < \infty$, which is $3/2$ for $N = 6$. The isothermal compressibility is $T\kappa_T = \delta n^2$.

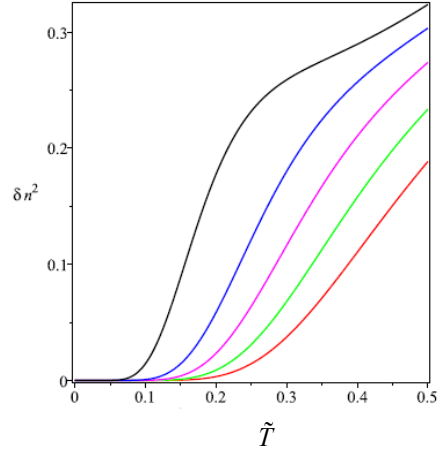

**Fig. 6.** On-site density fluctuations $\delta n^2$ versus scaled temperature $\tilde{T}$ for $N = 6$ with $\tilde{\mu} = -1.0$ (red), $\tilde{\mu} = -0.8$ (green), $\tilde{\mu} = -0.6$ (magenta), $\tilde{\mu} = -0.4$ (blue), and $\tilde{\mu} = -0.1$ (black). The correlations vanish at $\tilde{T} = 0$ with an essential singularity and vanish also for $|\tilde{\mu}| = \infty$ for $\tilde{T} > 0$. The isothermal compressibility is $T\kappa_T = \delta n^2$.

Next, we consider the correlation between two different sites. Correlation between sites is the result that the number of sites is equal to the maximum number possible of microstates at each site, viz., $\sum_{i=0}^{N} \hat{n}_i = \hat{N}$ Consider the correlations between sites, for instance, between sites 1 and 2, one obtains

$$C_{12} = \langle \hat{n}_1 \hat{n}_2 \rangle - \langle \hat{n}_1 \rangle \langle \hat{n}_2 \rangle = -\langle \hat{n}_1^2 \rangle + \langle \hat{n}_1 \rangle^2 - \sum_{j=3}^{N}[\langle \hat{n}_1 \hat{n}_j \rangle - \langle \hat{n}_1 \rangle \langle \hat{n}_j \rangle], \qquad (12)$$

for given $N$. In general, one has the sum-rule

$$\sum_{j=1}^{N} C_{ij} = 0 \qquad (i = 1, 2, \cdots N). \qquad (13)$$

It is interesting that for $N = 2$, one obtains from (12)-(13)

$$C_{12} = -C_{11} = -C_{22}. \qquad (14)$$

Therefore, the density correlation between the two lattice sites is the anti on-site density fluctuations of either lattice site.

## 7. Energy and entropy

The energy $E$ and the entropy $S$ follow from the grand canonical partition function $\mathscr{Z}$

$$E = \tilde{T}^2 \frac{\partial \ln \mathscr{Z}}{\partial \tilde{T}} + \tilde{\mu}\tilde{T}\frac{\partial \ln \mathscr{Z}}{\partial \tilde{\mu}} \quad \text{and} \quad S = \ln \mathscr{Z} + \tilde{T}\frac{\partial \ln \mathscr{Z}}{\partial \tilde{T}}, \qquad (15)$$

where we have set $k_B = 1$. We shall consider, in evaluating both the energy and the entropy, the simplest case of a lattice, viz., $N = 2$ and $\mu \equiv \mu_1 = -\mu_2$. Also, we will consider the case $N = 6$ with $\mu \equiv \mu_1 = \mu_2 = \cdots = \mu_6$. The former choice allows for the cases that one lattice site has two microstates while the other has zero microstates when $\mu \to \pm\infty$. Fig.7 shows the total energy as a function of the renormalized and scaled chemical potential $\tilde{\mu}$ for different values of the temperature $\tilde{T}$. Note the discontinuities in the ground state energy at $\tilde{\mu} = 0.5, 1.5$ (red) for $\tilde{T} \approx 0$. The latter singularities also appear in the entropy as seen in Fig.8 (red) for $\tilde{T} \approx 0$. Finally, in Fig.9, the entropy is plotted versus $\tilde{T}$. It is quite interesting that as $\tilde{T} \to \infty$, the entropy approaches the value $4\ln(2) = 2.7725...$ Recall that the maximum von Neumann entropy of a two-qubit entangled state is $\ln(2)$. Now we have four possible states in our two-site lattice and so in the high temperature limit the entropy reaches the value of $4\ln(2)$. In the general case of $N$ microstates per lattice site, the entropy approaches the value $N^2\ln 2$ as the temperature approaches infinity since there are $N^2$ qubits formed

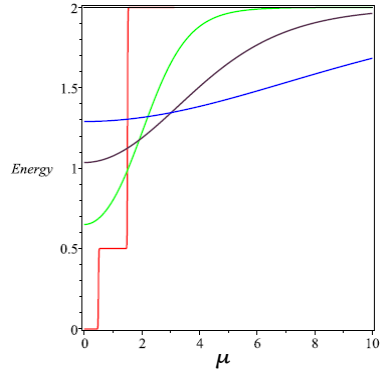

between two lattice sites with $N$ possible microstates in each lattice site.

**Fig. 7.** Energy versus renormalized and scaled $\tilde{\mu}$ for $N = 2$ with $\mu \equiv \mu_1 = -\mu_2$ and $\tilde{T} = 0.006$ (red), $\tilde{T} = 0.8$ (green), $\tilde{T} = 2.0$ (violet), $\tilde{T} = 5.0$ (blue), and $\tilde{T} = \infty$ (black), Energy = 2.

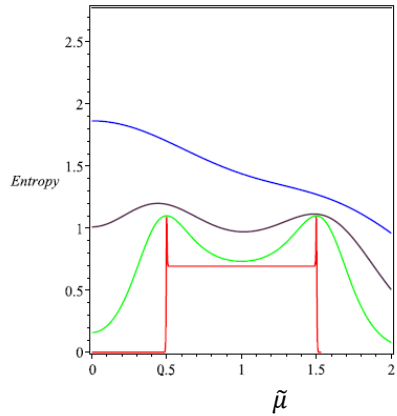

**Fig. 8.** Entropy versus renormalized and scaled $\tilde{\mu}$ for $N = 2$ with $\mu \equiv \mu_1 = -\mu_2$ and $\tilde{T} = 0.0015$ (red), $\tilde{T} = 0.1$ (green), $\tilde{T} = 0.2$ (violet), $\tilde{T} = 0.4$ (blue), $\tilde{T} = \infty$ (black) with Entropy = $4\ln(2) = 2.7725....$

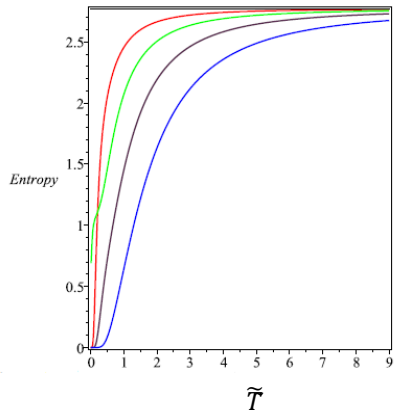

**Fig. 9.** Entropy versus scaled temperature $\tilde{T}$ for $N = 2$ with $\mu \equiv \mu_1 = -\mu_2$ and $\tilde{\mu} = 0.05$ (red), $\tilde{\mu} = 1.4$ (green), $\tilde{\mu} = 2.5$ (orange), $\tilde{\mu} = 4.0$ (blue), $\tilde{\mu} = \infty$ (black). Entropy $\to 4\ln(2) = 2.7725....$ as $\tilde{T} \to \infty$ for arbitrary $\tilde{\mu}$.

The plot in Fig.10 represents the energy for $N = 6$ lattice sites. Note the discontinues in the energy for $\tilde{\mu} = 1/2, 3/2, 5/2, 7/2, 9/2, 11/2$ for $\tilde{T} = 0$. All the different lattice sites with different chemical potentials have the same singularities. The energy approaches 108 in the limit $\tilde{\mu} \to \infty$.

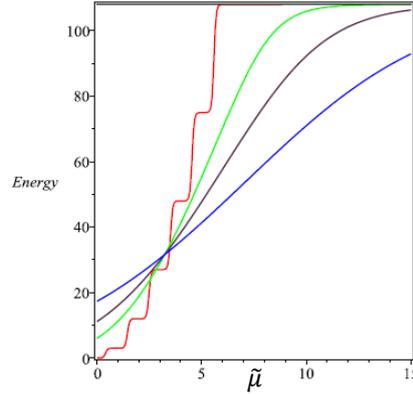

**Fig. 10.** Energy versus renormalized and scaled $\tilde{\mu}$ for $N = 6$ lattice sites with $\tilde{T} = 0.05$ (red), $\tilde{T} = 1.0$ (green), $\tilde{T} = 2.0$ (violet), $\tilde{T} = 4.0$ (blue), $\tilde{T} = \infty$ (black), where energy = 108. At $\tilde{T} = 0$, the energy = 0 for $0 < \tilde{\mu} < 1/2$, energy = 3.0 for $1/2 < \tilde{\mu} < 3/2$, energy = 12.0 for $3/2 < \tilde{\mu} < 5/2$, energy = 27.0 for $5/2 < \tilde{\mu} < 7/2$, energy = 48.0 for $7/2 < \tilde{\mu} < 9/2$, energy = 75.0 for $9/2 < \tilde{\mu} < 11/2$, and energy = 108 for $11/2 < \mu$.

The entropy is shown in Fig.11 as a function of $\tilde{\mu}$ with $\mu \equiv \mu_1 = \mu_2 = \cdots = \mu_6$. Note that the entropy has the same singularities as the energy for $\tilde{T} \approx 0$.

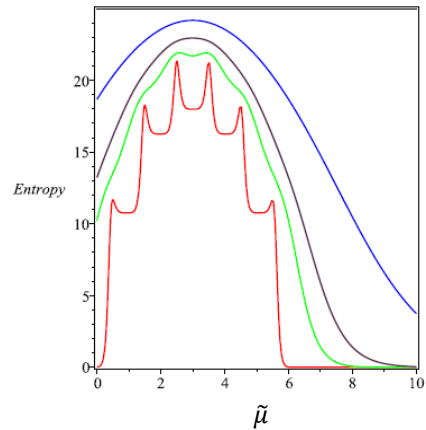

**Fig. 11.** Entropy versus renormalized and scaled $\tilde{\mu}$ for $N = 6$ lattice sites all with the same chemical potential $\tilde{\mu}$. $\tilde{T} = 0.05$ (red), $\tilde{T} = 0.3$ (green), $\tilde{T} = 0.5$ (violet), $\tilde{T} = 1.2$ (blue), $\tilde{T} = \infty$ (black), where entropy = $36\ln(2) = 24.95\ldots$

The entropy is plotted as a function of $\tilde{T}$ in Fig.12, where all the lattice sites have the same chemical potential, viz., $\mu \equiv \mu_1 = \mu_2 = \cdots = \mu_6$. The entropy exhibits a ground-state phase transition ($\tilde{T} = 0$) and the limit of the entropy as $\tilde{T} \to 0$ depends on the value of $\tilde{\mu}$. The latter is illustrated in Fig. 13.

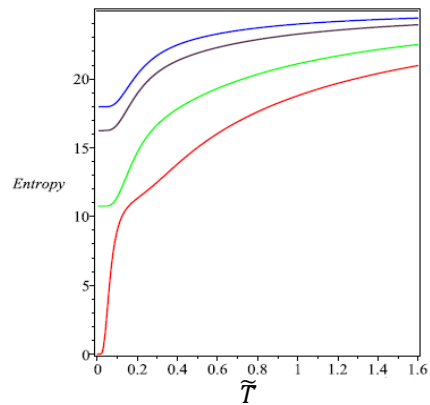

**Fig. 12.** Entropy versus scaled temperature $\tilde{T}$ for $N = 6$ lattice sites all with the same chemical potential $\tilde{\mu}$. $\tilde{\mu} = 0.3$ (red), $\tilde{\mu} = 1.0$ (green), $\tilde{\mu} = 2.0$ (violet), $\tilde{\mu} = 3.0$ (blue). Entropy $\rightarrow 36\ln(2) = 24.95\ldots$ (black) as $\tilde{T} \rightarrow \infty$.

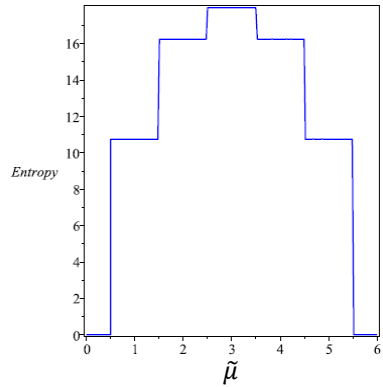

**Fig. 13.** Zero temperature behavior of the entropy versus renormalized and scaled $\tilde{\mu}$ for $N = 6$ lattice sites all with the same chemical potential $\tilde{\mu}$ for $\tilde{T} = 0.00001$. Entropy = 0 for $0 < \tilde{\mu} < 1/2$ and $11/2 < \tilde{\mu}$, entropy=10.75... for $1/2 < \tilde{\mu} < 3/2$ and $9/2 < \tilde{\mu} < 11/2$, entropy=16.24... for $3/2 < \tilde{\mu} < 5/2$ and $7/2 < \tilde{\mu} < 9/2$, entropy= 17.97... for $5/2 < \tilde{\mu} < 7/2$.

## Conclusions

We consider a modified Fermi-Hubbard model for a lattice with an equal number of microstates in each lattice site as the number of lattices sites. The modified model is soluble and we have calculated several thermodynamic variables in the grand canonical ensemble in terms of a renormalized and scaled chemical potential $\tilde{\mu}_i$ and a scaled temperature $\tilde{T}$. Our results are close to the original Fermi-Hubbard model for cases where $t \ll \mu$. The modified Fermi-Hubbard model exhibits a continuous phase transition (second order), which is reminiscent of the integer quantum Hall resistance. In addition, there is a ground-state, first-order phase transition where the value of the entropy as $\tilde{T} \rightarrow 0$ depends on the value of $\tilde{\mu}$.